\documentclass[ aps,%
floatfix,%
final,%
notitlepage,%
oneside,%
onecolumn,%
nobibnotes,%
nofootinbib,%
superscriptaddress,%
showpacs,%
]%
{revtex4}

\usepackage{epsfig}
\usepackage{amsfonts}
\usepackage{amsmath}
\usepackage{epsfig}
\usepackage{graphics}

\newcommand{\beq}{\begin{eqnarray}}\newcommand{\eeq}{\end{eqnarray}}
\newcommand{\beqa}{\begin{eqnarray*}}\newcommand{\eeqa}{\end{eqnarray*}}

\begin{document}

\title{ Exclusive $C=+$ charmonium  production in $e^+ e^- \to H+\gamma$ at B-factories 
\\ within light cone formalism. }
\author{V.V. Braguta}

\email{braguta@mail.ru}

\affiliation{Institute for High Energy Physics, Protvino, Russia}

\begin{abstract}
In this paper the cross sections of the processes 
$e^+ e^- \to H+\gamma, H=\eta_c, \eta_c', \chi_{c0}, \chi_{c1}, \chi_{c2}$ are calculated. 
The calculation is carried out at the leading twist approximation of light cone
formalism. Within this approach the leading logarithmic radiative and relativistic corrections
to the amplitudes are resumed. For the processes $e^+ e^- \to \eta_c, \eta_c'+\gamma$
one loop radiative corrections are taken into account. It is also shown 
that one loop leading logarithmic radiative corrections calculated within light cone
formalism for the processes under study coincide with that obtained by direct 
calculations of one loop diagrams within nonrelativistic QCD.
\end{abstract}


\maketitle

\newcommand{\ins}[1]{\underline{#1}}
\newcommand{\subs}[2]{\underline{#2}}
\vspace*{-1.cm}
\section{Introduction}

Theoretical approach to the description of hard exclusive processes, which is called light 
cone formalism (LC), is based on 
the factorization theorem \cite{Lepage:1980fj, Chernyak:1983ej}. Within this theorem the amplitude of hard exclusive process
can be separated into two parts. The first part is partons production at very small 
distances, which can be treated within perturbative QCD. The second part is 
the hardronization of the partons at large distances. For hard exclusive processes it can be 
parameterized by process independent distribution amplitudes (DA).  

The production of the charmonium meson $H$ in the process $e^+ e^- \to H+\gamma$ at B-factories, 
is the simplest example of hard exclusive process. One can assume that the energy at 
which B-factories operate is sufficiently large so that it is possible to apply 
LC. Another approach to the calculation of the cross section 
of this process is nonrelativistic QCD (NRQCD) \cite{Bodwin:1994jh}. This approach is based 
on the assumption that relative velocity of quark-antiquark pair in charmonia  is small 
parameter in which the amplitude of charmonium production can be expanded.  
LC has two very important advantages in comparison to the NRQCD. The first 
advantage is that LC formalism can be applied for light or heavy mesons if DA 
of this meson is known. From, NRQCD perspective this means that LC resums whole 
series of relativistic corrections to the amplitude under study. For NRQCD relativistic 
corrections are very important especially for the production of  exited charmonia 
mesons. The second advantage is that within LC one can resum leading logarithmic 
radiative corrections to the amplitude in all loops. The main disadvantage of 
LC is that within this formalism it is rather difficult to control power corrections
to the amplitude. 

Within NRQCD the process $e^+ e^- \to H+\gamma$ was considered in papers \cite{Chung:2008km, Li:2009ki,  Sang:2009jc}. 
In paper \cite{Chung:2008km} this process was considered at the leading order approximation in relative velocity
and strong coupling constant. The authors of paper \cite{Li:2009ki} took into 
account one loop radiative corrections. In addition to the radiative corrections
the first order relativistic corrections to the process $e^+ e^- \to \eta_c+\gamma$
were considered in paper \cite{Sang:2009jc}.

The only process considered within LC is 
$e^+ e^- \to \eta_c+\gamma$ \cite{Jia:2008ep, shifman}. The main drawback 
of these papers is that  the authors used very simple model of DA of the $\eta_c$ meson, 
which doesn't take into account relativistic motion in this meson. 
Recently, the leading twist DAs of charmonia mesons have become the object of intensive 
study
\cite{Bodwin:2006dm, Ma:2006hc, Braguta:2006wr, Braguta:2007fh, 
Braguta:2007tq, Choi:2007ze, Bell:2008er, Braguta:2008qe, Hwang:2009cu}.
The study of these DAs allowed one to build some models for charmonia DAs, 
that can be used in the calculation of different exclusive processes.

In this paper the leading twist processes $e^+ e^- \to H+\gamma$ will be considered. 
Using helicity selection rules \cite{Chernyak:1977fk,Chernyak2,Chernyak3} it is not difficult to show that at the leading 
twist accuracy the mesons with longitudinal polarization and  the following quantum numbers $H=^1S_0, ^3P_1, ^3P_2, ^3P_3$ can be produced. 
So, in this paper the following processes will be considered: $e^+ e^- \to H+\gamma, H=\eta_c, \eta_c', 
\chi_{c0}, \chi_{c1}, \chi_{c2}$. To calculate the cross sections 
of these processes the model of DAs proposed in papers 
\cite{Braguta:2006wr, Braguta:2007fh, Braguta:2007tq, Braguta:2008qe} will be used.

This paper is organized as follows. In the next section the amplitudes of the 
processes under consideration will be derived. Numerical results and the 
discussion of these results will be given in the last section of this paper.

\section{The amplitude of the process $e^+ e^- \to H+\gamma$.}

\begin{figure}
\begin{centering}
\includegraphics[scale=0.8]{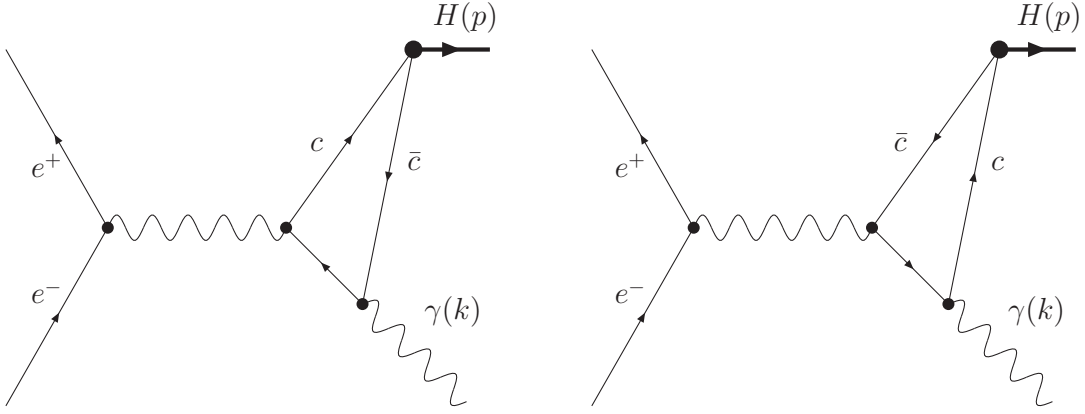}
\par\end{centering}
\caption{The diagrams that contribute to the processes $e^+ e^- \to H+\gamma, H=\eta_c, 
\eta_c', \chi_{c0}, \chi_{c1}, \chi_{c2}$ at the leading order approximation. 
in the strong coupling constant}
\label{fig1}
\end{figure}

In this section the leading twist approximation for the amplitude of the processes 
$e^+ e^- \to H+\gamma, H=\eta_c, \eta_c', \chi_{c0}, \chi_{c1}, \chi_{c2}$ will be derived. 
The diagrams that contribute to the processes  at the leading order approximation 
in the strong coupling constant are shown in Fig. \ref{fig1}
As was noted in the introduction, selection rules tell us that at the leading 
twist accuracy all produced mesons are longitudinally polarized. So, the
polarization vectors of these mesons are proportional to the momentum of these 
mesons. 

To calculate the amplitudes and the cross sections of the processes involved one needs 
the expressions for the following matrix element of the electromagnetic current 
$J_{\mu}(0)$: $~\langle H(p) \gamma(k)| J_{\mu}(0)|0 \rangle$. For the production 
of the longitudinally polarized $\eta_c, \eta_c', \chi_{c1}$ mesons it can parameterized as follows
\beq
\langle H(p) \gamma(k)| J_{\mu}(0)|0 \rangle = F_H ~e_{\mu \nu \alpha \beta} \epsilon^{\nu} p^{\alpha} k^{\beta},
\label{J1}
\eeq
where $\epsilon^{\nu}$ is the polarization vector of the final photon. It 
causes no difficulties to find the expression for the 
formfactor $F_{H=\eta_c, \eta_c', \chi_{c1}}$ at the leading
twist approximation 
\beq
F_{\eta_c, \eta_c', \chi_{c1}} = \frac {16 \pi \alpha Q_c^2 f_{\eta_c, \eta_c', \chi_{c1}}} {s}
\int_{-1}^{1} d \xi \frac {\phi_{\eta_c, \eta_c', \chi_{c1}}( \xi, \mu)} {(1-\xi^2)}, 
\label{HJ1}
\eeq
where $Q_c$ is the charge of $c$-quark, the definitions of the constants $f_{\eta_c, \eta_c', \chi_{c1}}$ 
and the DAs $\phi_{\eta_c, \eta_c', \chi_{c1}}( \xi, \mu)$ can be found in the Appendix, 
$\xi$ is the fraction of relative momentum of the whole meson carried by quark-antiquark pair, 
$\mu$ is the characteristic scale of the process, $s=(p+k)^2$.

The expression for the production amplitude
of the longitudinally polarized $ \chi_{c0}, \chi_{c2}$ mesons can be written 
in the following form
\beq
\langle H(p) \gamma(k)| J_{\mu}(0)|0 \rangle = F_H~\biggl ( 
(\epsilon q) k_{\mu} - (kq) \epsilon_{\mu}
\biggr ),
\label{J2}
\eeq
where $q=k+p$. The other designations are the same as were used in equation (\ref{J1}).
The expression for the formfactor $F_{H=\chi_{c0}, \chi_{c2}}$ has the form
\beq
F_{\chi_{c0}, \chi_{c2} } = \frac {16 \pi \alpha Q_c^2 f_{ \chi_{c0}, \chi_{c2} }} {s}
\int_{-1}^{1} d \xi \frac {\xi ~\phi_{\chi_{c0}, \chi_{c2}}( \xi, \mu)} {(1-\xi^2)}, 
\label{HJ2}
\eeq
The constants $f_{\chi_{c0}, \chi_{c2}}$ 
and the DAs $\phi_{\chi_{c0}, \chi_{c2}} ( \xi, \mu)$ can be found in the Appendix.

The cross section of the processes can be written in the following form
\beq
\sigma_{H}= \frac {\alpha} {24} 
F^2_{H} \biggl ( 1- \frac {M_H^2} {s} \biggr ),
\label{cr}
\eeq
It should be noted that the matrix elements of the processes under study were taken at the 
leading order approximation in ${M_H^2}/{s}$. The factor 
 $1-  {M_H^2}/ {s}$ in the cross section appeared due to the phase space of the final particles. 

The expression of the formfactor $F_H$ depend on the DAs $\phi_H(\xi, \mu)$ of the charmonia
mesons. If infinitely narrow distribution amplitudes 
$\phi_{\eta_c, \eta_c', \chi_{c1}}(\xi, \mu)=\delta(\xi),~~ 
\phi_{\chi_{c0}, \chi_{c2} }(\xi, \mu)=-\delta'(\xi) $ are substituted 
to formulas (\ref{HJ1}), (\ref{HJ2}), than NRQCD results for the amplitude will be reproduced  
\cite{Chung:2008km}. 
If real distribution amplitudes $\phi_H(\xi, \mu)$ are taken at the scale 
$\mu \sim m_c$, than formulas (\ref{HJ1}), (\ref{HJ2}) will resum the relativistic corrections 
to the cross section up to $O(1/s^3)$ terms. 
To resum  the relativistic and leading logarithmic radiative corrections simultaneously
one must take the distribution amplitudes $\phi_H(\xi, \mu)$ at the 
characteristic scale of the process $\mu \sim \sqrt s$. The calculation  
of the cross sections  will be done at the scale $\mu=\sqrt s /2$.

It is interesting to note that it is possible find the leading logarithmic 
radiative corrections at one loop level using formulas (\ref{HJ1}), (\ref{HJ2}) without 
calculation of one loop diagrams. Applying the approach 
proposed in paper \cite{Jia:2008ep} one gets the results
\beq
F_{\eta_c, \eta_c'}=\frac {16 \pi \alpha Q_c^2 } {s} \sqrt{ \frac {\langle O \rangle_S} {m_c} }
\biggl (1+ C_f \frac {\alpha_s( s ) } {4 \pi} \log {\biggl (\frac {\mu^2} {\mu_0^2}} \biggr )
\bigl ( 3-2 \log 2 \bigr ) \biggr ),   \nonumber \\
F_{\chi_{c0}}=\frac {16 \pi \alpha Q_c^2 } {s} \sqrt{ \frac {\langle O \rangle_P} {3 m_c^3} }
\biggl (1+ C_f \frac {\alpha_s( s ) } {4 \pi} \log {\biggl (\frac {\mu^2} {\mu_0^2}} \biggr )
\bigl ( 1- 2 \log 2  \bigr  ) \biggr ), \nonumber \\
F_{\chi_{c1}}=\frac {16 \pi \alpha Q_c^2 } {s} \sqrt{ \frac {2 \langle O \rangle_P} {m_c^3} }
\biggl (1+ C_f \frac {\alpha_s( s ) } {4 \pi} \log {\biggl (\frac {\mu^2} {\mu_0^2}} \biggr )
\bigl ( 3-2 \log 2 \bigr ) \biggr ),  \nonumber \\
F_{\chi_{c2}}=\frac {16 \pi \alpha Q_c^2 } {s}  \sqrt{ \frac {2 \langle O \rangle_P} {3 m_c^3} }
\biggl (1+ C_f \frac {\alpha_s( s ) } {4 \pi} \log {\biggl (\frac {\mu^2} {\mu_0^2}} \biggr )
\bigl ( 1- 2 \log 2  \bigr ) \biggr ),
\label{LL}
\eeq
where $m_c$ is the pole mass of the $c$-quark, the definition of the NRQCD matrix elements 
$\langle O \rangle_S, \langle O \rangle_P$ can found in paper \cite{Bodwin:1994jh}, $C_f=4/3$. 
Note that in the above equations it was assumed that renormalization group evolution 
of the DAs begins at the scale $\mu_0 \sim m_c$ and ends at the scale $\mu \sim \sqrt s$.
At the scale $\mu_0 \sim m_c$ the DAs are $\phi_{\eta_c, \eta_c', \chi_{c1}}(\xi, \mu)=\delta(\xi),~~ 
\phi_{\chi_{c0}, \chi_{c2} }(\xi, \mu)=-\delta'(\xi) $. Leading order NRQCD results 
(\ref{LL}) coincide with the results obtained in paper \cite{Chung:2008km}. 
The one loop leading logarithmic radiative corrections for the $F_{\eta_c}$ coincide with the result
of paper \cite{Jia:2008ep}. The one loop leading logarithmic radiative corrections for the 
$F_{\eta_c}, F_{\chi_{c0}}, F_{\chi_{c1}}, F_{\chi_{c2}}$ coincide with the results 
of paper \cite{Sang:2009jc}. 

The result (\ref{HJ1}) for the production of the pseudoscalar mesons $\eta_c, \eta_c'$ 
can be improved since there exists expression for the one loop radiative correction
to this amplitude \cite{Braaten:1982yp, Kadantseva:1985kb}. This expression can be written as follows \cite{Braaten:1982yp}
\beq
F_{\eta_c, \eta_c'} = \frac {16 \pi \alpha Q_c^2 f_{\eta_c, \eta_c'}} {s}
\int_{-1}^{1} d \xi \frac {\phi_{\eta_c, \eta_c'}( \xi, \mu)} {(1+\xi)} 
\biggl [ 
1+ C_f \frac {\alpha_s (s)} {4 \pi} \biggl ( &&
\log^2 {\biggl ( \frac {1+\xi} 2 \biggr ) }  
- \frac {1+\xi} {1-\xi} \log { \biggl ( \frac {1+\xi} 2 \biggr ) } \nonumber  \\ 
&& - 9 +
\biggl ( 
3+2 \log {\biggl ( \frac {1+\xi} 2 \biggl ) }
\biggr ) \log {\biggl ( \frac s {\mu^2} \biggl ) }
\biggr )
\biggr ], 
\label{rad}
\eeq
In the above expressions it is assumed that the DAs $\phi_{\eta_c, \eta_c'}$
are $\xi$ even. 

It is instructive to take the limit of zero relative velocity of quark-antiquark pair 
and compare it to the NRQCD result \cite{Sang:2009jc}. At leading order approximation 
in relative velocity expression (\ref{rad}) becomes
\beq
F_{\eta_c, \eta_c'}=\frac {16 \pi \alpha Q_c^2 } {s} \sqrt{ \frac {\langle O \rangle_S} {m_c} }
\biggl [1+ && C_f \frac {\alpha_s( s ) } {4 \pi} \log {\biggl (\frac {\mu^2} {\mu_0^2}} \biggr )
\biggl ( 3-2 \log 2 \biggr ) 
 \nonumber \\ && + C_f \frac {\alpha_s( s ) } {4 \pi}
\biggl (
\log^2 2 +\log 2 - 9 + \log {\biggl (\frac {s} {\mu^2}} \biggr )
\bigl ( 3-2 \log 2 \bigr )
\biggr )
\biggr ].
\label{nrrad}
\eeq
The second term in equation (\ref{nrrad}) is due to renormalization 
group resummation of the leading logarithms in the DA. The last term is 
one loop radiative corrections to the hard part of the amplitude. 
The factorization scale $\mu$ separates long distance dynamic of 
the chamonium meson parameterized by DA from the small distance effects
parameterized in the hard part of the amplitude. It is seen that
$\mu$ dependence is canceled in the final answer, as it should be.

The authors of paper \cite{Sang:2009jc} obtained the following 
NRQCD expression for  equation (\ref{nrrad})  
\beq
F_{\eta_c, \eta_c'}=\frac {16 \pi \alpha Q_c^2 } {s} \sqrt{ \frac {\langle O \rangle_S} {m_c} }
\biggl [1+ C_f \frac {\alpha_s( s ) } {4 \pi}
\biggl (
\log^2 2 +3 \log 2 - 9 - \frac {\pi^2} 3+ \log {\biggl (\frac {s} {m_c^2}} \biggr ) 
\bigl ( 3-2 \log 2 \bigr )
\biggr )
\biggr ].
\label{nr1rad}
\eeq
It is seen that this expression is very similar to (\ref{nrrad}). Moreover 
one has one free parameter $\mu_0$ in expression  (\ref{nrrad}), which 
can be used to adjust  (\ref{nrrad}) to  (\ref{nr1rad}). However, expressions
(\ref{nrrad}) and  (\ref{nr1rad}) seem to be a little bit different.

\section{Numerical results and discussion.}

To obtain numerical results for the cross sections of the processes 
under study the following numerical parameters are needed.

In this paper we are going to use the models of the charmonia DAs proposed in papers
\cite{Braguta:2006wr, Braguta:2007fh, Braguta:2007tq, Braguta:2008qe}.
For the strong coupling constant we use one-loop expression
\begin{eqnarray*}
  \alpha_s(\mu) &=& \frac{4\pi}{b_0 \ln(\mu^2/\Lambda_\mathrm{QCD}^2)},
\end{eqnarray*}
where $b_0=25/3$ and $\Lambda_\mathrm{QCD}=0.2$ GeV.

In the calculation the following values of the constants $f_H$ will be used \cite{Braguta:2009df}
\begin{eqnarray}
f_{\eta_c} & = & 0.373 \pm 0.064 \,\mathrm{GeV}, \nonumber \\
f_{\eta_c'} &=& 0.261 \pm 0.077 \,\mathrm{GeV},\nonumber \\
f_{\chi_{c0}}(M_{J/\Psi}) & = & 0.093 \pm 0.017\,\mathrm{GeV},\nonumber \\
f_{\chi_{c1}} & = & 0.272 \pm 0.048 \,\mathrm{GeV}, \nonumber \\
f_{\chi_{c2}}(M_{J/\Psi}) & = & 0.131 \pm 0.023 \,\mathrm{GeV}
\label{const_values}
\end{eqnarray}
The values of the constants $f_{\eta_c}, f_{\eta_c'}$ were
calculated in paper \cite{Braguta:2008tg}.  The values of the constants of the $P$-wave charmonia mesons
can be found in paper \cite{Braguta:2008qe}. It should be noted that the constants
$ f_{\chi_{c0}}, f_{\chi_{c2}}$ depend on the renormalization scale. As it is seen from formulas (\ref{const_values})
these constants are defined at the scale $\mu=M_{J/\Psi}$. The anomalous dimensions of these constants,
which govern the evolution, can be found in paper \cite{Braguta:2008qe}.

\begin{table}
$$\begin{array}{|c|c|c|c|c|}
\hline
 H  & \sigma (e^+ e^- \to H+\gamma) (\mbox{fb})
   &  \sigma (e^+ e^- \to H+\gamma) (\mbox{fb})
 & \sigma (e^+ e^- \to H+\gamma) (\mbox{fb})    & \sigma (e^+ e^- \to H+\gamma) (\mbox{fb})  \\
  & \mbox{This work}  & \mbox{\cite{Chung:2008km}} & \mbox{\cite{Li:2009ki}} & \mbox{\cite{Sang:2009jc}}  \\
\hline
\eta_c& 41.6 \pm 14.1 & 82.0^{+21.4}_{-19.8} & 42.5-53.7  & 68.0^{+22.2}_{-20.3}  \\
\hline
\eta_c'& 24.2 \pm  14.5 & 49.2^{+9.4}_{-7.4} & 27.7-35.1 &  42.6^{+10.9}_{-8.8} \\
\hline
\chi_{c0}& 6.1 \pm 3.9 & 1.3^{+0.2}_{-0.2}  & 1.53-2.48   &  1.36^{+0.26}_{-0.26} \\
\hline
\chi_{c1}& 24.2 \pm 13.3  & 13.7^{+3.4}_{-3.1}  & 11.1-17.7 & 10.9^{+3.7}_{-3.4}  \\
\hline
\chi_{c2}& 12.0 \pm  17.4 & 5.3^{+1.6}_{-1.3} & 1.65-3.53 & 1.95^{+1.85}_{-1.56} \\
\hline
\end{array}$$
\label{tab}
\caption{ The cross sections of the processes 
$e^+ e^- \to H+\gamma, H=\eta_c, \eta_c', \chi_{c0}, \chi_{c1}, \chi_{c2}$. 
Second column contains the results obtained in this paper.  
In the third, fourth and fifth columns the results   
obtained in papers \cite{Chung:2008km}, \cite{Li:2009ki}, \cite{Sang:2009jc} are shown. 
}
\end{table}

There are different sources of uncertainty to the results obtained in this paper. The most important 
uncertainties can be divided into the following groups:

{\bf 1.} {\it The uncertainty in the models of the distribution amplitudes $\phi_H (x, \mu)$}, 
which can be modeled by the variation of the parameters of these models .
The calculation shows that this source of uncertainty is not greater than 10\%. So, it 
is not very important and it will be ignored.

{\bf 2.} {\it The uncertainty due to  radiative corrections}. 
In the approach applied in this paper the leading logarithmic radiative corrections due to the 
evolution of the DAs and the strong coupling constant were resummed. For the processes
$e^+e^- \to \eta_c, \eta_c'+\gamma$ one loop radiative corrections were
taken into account. So, for last two processes radiative corrections 
are not very important and they will be ignored. As to the 
other processes considered in this paper radiative corrections to the results 
can be estimated as $\alpha_s(s) \sim 20 \%$.

{\bf 3.} {\it The uncertainty due to the power corrections.} This uncertainty is determined 
by the next-to-leading order contribution in the $1/s$ expansion. One can estimate these 
corrections using the leading order NRQCD predictions \cite{Chung:2008km}, as it was discussed 
in paper \cite{Braguta:2009df}. Thus, for 
the processes $e^+e^- \to \eta_c, \eta_c', \chi_{c0}, \chi_{c1}, \chi_{c2}+ \gamma$ the
errors due to this source of uncertainty  are $\sim 3 \%, 6 \%, 50 \%,  37 \%, 60 \%$
correspondingly.

{\bf 4.} {\it The uncertainty in the values of constants (\ref{const_values}).}  The calculations 
show that, for the processes 
$e^+e^- \to \eta_c, \eta_c', \chi_{c0}, \chi_{c1}, \chi_{c2}+ \gamma$ the
errors due to this source of uncertainties  are 
$\sim 34 \%, 60 \%, 35 \%,  35 \%, 35 \%$ correspondingly. 

Adding all these uncertainties in quadrature one gets the total errors of the calculation.

The results of the calculation are presented in Table \ref{tab}. Second column 
contains the results obtained in this paper.  In the third, fourth and fifth columns the results   
obtained in papers \cite{Chung:2008km}, \cite{Li:2009ki}, \cite{Sang:2009jc} are shown.
It is seen that the results obtained in this paper are in reasonable agreement 
with the results obtained within NRQCD.

\begin{acknowledgments}
This work was partially supported by Russian Foundation of Basic Research under grant 08-02-00661, grant 09-01-12123, grant 10-02-00061, Leading Scientific Schools grant NSh-6260.2010.2 
 and president grant MK-140.2009.2.

\end{acknowledgments}

\appendix

\section{Distribution amplitudes.}

The leading twist distribution amplitudes needed in the calculation can be defined as
follows:

{\bf for the pseudoscalar mesons $P=\eta_c, \eta_c'$:}
\begin{eqnarray*}
\left\langle P(p)\left|
\bar{Q}^i_{\alpha}(z) [z,-z]Q^j_{\beta}(-z)\right|0\right\rangle  & = &
(\hat p \gamma_5)_{\beta \alpha} \frac {f_P}{4} \frac{\delta_{ij}}{3}
\int\limits _{-1}^{1}d\xi e^{i\xi(pz)}\phi_{\eta_c}(\xi;\mu),
\end{eqnarray*}

{\bf  for the $\chi_{c0}$-meson:}
\begin{eqnarray*}
\left\langle \chi_{c0}(p)\left|\bar{Q}^i_{\alpha} (z) [z,-z] Q^j_{\beta}(-z)\right|0\right\rangle  & = &
(\hat p)_{\beta \alpha}  \frac{f_{\chi_0}}{4}  \frac{\delta_{ij}}{3}
\int\limits _{-1}^{1}d\xi e^{i\xi(pz)}\phi_{\chi_0}(\xi;\mu)
\end{eqnarray*}

{\bf  for the $\chi_{c1}$-meson:}
\begin{eqnarray*}
\left\langle \chi_{c1}(p,\epsilon_{\lambda=0})\left|\bar{Q}^i_{\alpha} (z) [z,-z] Q^j_{\beta}(-z)\right|0\right\rangle  & = &
(\hat p \gamma_5)_{\beta \alpha} \frac{f_{\chi_1}}{4}  \frac{\delta_{ij}}{3}
\int\limits _{-1}^{1}d\xi e^{i\xi(pz)}\phi_{\chi_1}(\xi;\mu),
\end{eqnarray*}

{\bf  for the $\chi_{c2}$-meson:}
\begin{eqnarray*}
\left\langle \chi_{c2}(p,\epsilon_{\lambda=0})\left|\bar{Q}^i_{\alpha} (z) [z,-z] Q^j_{\beta}(-z)\right|0\right\rangle  & = &
(\hat p)_{\beta \alpha}  \frac{f_{\chi_2}}{4} \frac{\delta_{ij}}{3}
\int\limits _{-1}^{1}d\xi e^{i\xi(pz)}\phi_{\chi_2}(\xi;\mu),
\end{eqnarray*}
The factor $[z,-z]$, that
makes the above matrix elements
gauge invariant, is defined as
\begin{eqnarray*}
[z, -z] = P \exp[i g \int_{-z}^z d x^{\mu} A_{\mu} (x) ].
\end{eqnarray*}

It is not difficult to show  that the functions $\phi_{\eta_c}(\xi)$ and
 $\phi_{\chi_1}(\xi)$ 
are $\xi$-even. The normalization condition for these functions is
\begin{eqnarray*}
\int\limits _{-1}^{1}\phi(\xi)d\xi & = & 1.\label{eq:normE}\end{eqnarray*}
 The functions $\phi_{\chi_0}(\xi)$ and $\phi_{\chi_2} (\xi)$
 are $\xi$-odd and normalized according to \begin{eqnarray*}
\int\limits _{-1}^{1}\xi\phi(\xi)d\xi & = & 1.\end{eqnarray*}


\begin{thebibliography}{**}



\bibitem{Lepage:1980fj}
  G.~P.~Lepage and S.~J.~Brodsky,
  Phys.\ Rev.\ D {\bf 22}, 2157 (1980).

\bibitem{Chernyak:1983ej}
  V.~L.~Chernyak and A.~R.~Zhitnitsky,
  Phys.\ Rept.\  {\bf 112}, 173 (1984).

\bibitem{Bodwin:1994jh}
  G.~T.~Bodwin, E.~Braaten and G.~P.~Lepage,
  Phys.\ Rev.\ D {\bf 51}, 1125 (1995)
  [Erratum-ibid.\ D {\bf 55}, 5853 (1997)]
  [arXiv:hep-ph/9407339].


\bibitem{Chung:2008km}
  H.~S.~Chung, J.~Lee and C.~Yu,
  Phys.\ Rev.\  D {\bf 78}, 074022 (2008)
  [arXiv:0808.1625 [hep-ph]].


\bibitem{Li:2009ki}
  D.~Li, Z.~G.~He and K.~T.~Chao,
  Phys.\ Rev.\  D {\bf 80}, 114014 (2009)
  [arXiv:0910.4155 [hep-ph]].

\bibitem{Sang:2009jc}
  W.~L.~Sang and Y.~Q.~Chen,
  Phys.\ Rev.\  D {\bf 81}, 034028 (2010)
  [arXiv:0910.4071 [hep-ph]].



\bibitem{Jia:2008ep}
  Y.~Jia and D.~Yang,
  Nucl.\ Phys.\  B {\bf 814}, 217 (2009)
  [arXiv:0812.1965 [hep-ph]].

\bibitem{shifman}
  M.A.~Shifman and M.I.~Vysotsky, Nucl. Phys. B {\bf 186}, 475, (1981).


\bibitem{Bodwin:2006dm}
  G.~T.~Bodwin, D.~Kang and J.~Lee,
  Phys.\ Rev.\  D {\bf 74}, 114028 (2006)
  [arXiv:hep-ph/0603185].


\bibitem{Ma:2006hc}
  J.~P.~Ma and Z.~G.~Si,
  Phys.\ Lett.\  B {\bf 647}, 419 (2007)
  [arXiv:hep-ph/0608221].


\bibitem{Braguta:2006wr}
  V.~V.~Braguta, A.~K.~Likhoded and A.~V.~Luchinsky,
  Phys.\ Lett.\  B {\bf 646}, 80 (2007)
  [arXiv:hep-ph/0611021].

\bibitem{Braguta:2007fh}
  V.~V.~Braguta,
  Phys.\ Rev.\  D {\bf 75}, 094016 (2007)
  [arXiv:hep-ph/0701234].



\bibitem{Braguta:2007tq}
  V.~V.~Braguta,
  Phys.\ Rev.\  D {\bf 77}, 034026 (2008)
  [arXiv:0709.3885 [hep-ph]].


\bibitem{Choi:2007ze}
  H.~M.~Choi and C.~R.~Ji,
  Phys.\ Rev.\  D {\bf 76}, 094010 (2007)
  [arXiv:0707.1173 [hep-ph]].



\bibitem{Bell:2008er}
  G.~Bell and T.~Feldmann,
  JHEP {\bf 0804}, 061 (2008)
  [arXiv:0802.2221 [hep-ph]].




\bibitem{Braguta:2008qe}
  V.~V.~Braguta, A.~K.~Likhoded and A.~V.~Luchinsky,
  Phys.\ Rev.\  D {\bf 79}, 074004 (2009)
  [arXiv:0810.3607 [hep-ph]].

\bibitem{Hwang:2009cu}
  C.~W.~Hwang,
  JHEP {\bf 0910}, 074 (2009)
  [arXiv:0906.4412 [hep-ph]].

\bibitem{Chernyak:1977fk} V.~L.~Chernyak, A.~R.~Zhitnitsky and
V.~G.~Serbo, 
 JETP Lett.\ \textbf{26}, 594 (1977) {[}Pisma Zh.\ Eksp.\ Teor.\ Fiz.\ \textbf{26},
760 (1977)]. 


\bibitem{Chernyak2} V.~L.~Chernyak and A.~R.~Zhitnitsky, 
 Sov.\ J.\ Nucl.\ Phys.\ \textbf{31}, 544 (1980) {[}Yad.\ Fiz.\ \textbf{31},
1053 (1980)]. 


\bibitem{Chernyak3} V.~L.~Chernyak and A.~R.~Zhitnitsky, 
 JETP Lett.\ \textbf{25}, 510 (1977) {[}Pisma Zh.\ Eksp.\ Teor.\ Fiz.\ \textbf{25},
544 (1977)]. 



\bibitem{Braaten:1982yp}
  E.~Braaten,
  Phys.\ Rev.\  D {\bf 28}, 524 (1983).

\bibitem{Kadantseva:1985kb}
  E.~P.~Kadantseva, S.~V.~Mikhailov and A.~V.~Radyushkin,
  Yad.\ Fiz.\  {\bf 44}, 507 (1986)
  [Sov.\ J.\ Nucl.\ Phys.\  {\bf 44}, 326 (1986)].

\bibitem{Braguta:2009df}
  V.~V.~Braguta, A.~K.~Likhoded and A.~V.~Luchinsky,
  Phys.\ Rev.\  D {\bf 80}, 094008 (2009)
  [arXiv:0902.0459 [hep-ph]].

\bibitem{Braguta:2008tg}
  V.~V.~Braguta,
  Phys.\ Rev.\  D {\bf 79}, 074018 (2009)
  [arXiv:0811.2640 [hep-ph]].






\end{thebibliography}
\end{document}